\documentclass[12pt]{article}

\usepackage{graphicx}

\usepackage{times}

\newenvironment{sciabstract}{%
\begin{quote} \bf}
{\end{quote}}

\newcounter{lastnote}

\title{The fastest unbound star in our Galaxy ejected by a thermonuclear supernova}

\author
{S. Geier$^{1,2}$, F. F\"urst$^{3}$, E. Ziegerer$^{2}$, T. Kupfer$^{4}$, U. Heber$^{2}$,\\
A. Irrgang$^{2}$, B. Wang$^{5}$, Z. Liu$^{5,6}$, Z. Han$^{5}$, B. Sesar$^{7,8}$, D. Levitan$^{7}$,\\
R. Kotak$^{9}$, E. Magnier$^{10}$, K. Smith$^{9}$, W. S. Burgett$^{10}$,\\ 
K. Chambers$^{10}$, H. Flewelling$^{10}$, N. Kaiser$^{10}$, R. Wainscoat$^{10}$, C. Waters$^{10}$\\
\\
\normalsize{$^{1}$European Southern Observatory, Karl-Schwarzschild-Str.~2,}\\ 
\normalsize{85748 Garching, Germany}\\
\normalsize{$^{2}$Dr.\,Karl Remeis-Observatory \& ECAP, Astronomical Institute,}\\ 
\normalsize{Friedrich-Alexander University Erlangen-Nuremberg, Sternwartstr.~7, 96049 Bamberg, Germany}\\
\normalsize{$^{3}$Space Radiation Lab, MC 290-17 Cahill, California Institute of Technology,}\\
\normalsize{1200 E. California Blvd, Pasadena, CA 91125, USA}\\
\normalsize{$^{4}$Department of Astrophysics/IMAPP, Radboud University Nijmegen,}\\ 
\normalsize{P.O. Box 9010, 6500 GL Nijmegen, The Netherlands}\\
\normalsize{$^{5}$Key Laboratory of the Structure and Evolution of Celestial Objects,}\\ 
\normalsize{Yunnan Observatories, Chinese Academy of Sciences, Kunming 650011, China}\\
\normalsize{$^{6}$Argelander-Institut f\"ur Astronomie,}\\
\normalsize{Universit\"at Bonn, Auf dem H\"ugel 71, 53121, Bonn, Germany}\\
\normalsize{$^{7}$Division of Physics, Mathematics, and Astronomy,}\\ 
\normalsize{California Institute of Technology, 1200 E. California Blvd, Pasadena, CA 91125, USA}\\ 
\normalsize{$^{8}$Max-Planck-Institut für Astronomie,}\\
\normalsize{K\"onigstuhl 17, 69117, Heidelberg, Germany}\\
\normalsize{$^{9}$Astrophysics Research Centre, School of Mathematics and Physics,}\\ 
\normalsize{Queen’s University Belfast, Belfast BT7 1NN, UK}\\
\normalsize{$^{10}$Institute for Astronomy, University of Hawaii at Manoa,}\\ 
\normalsize{Honolulu, HI 96822, USA}\\
\\
\normalsize{$^\ast$E-mail: sgeier@eso.org.}
}

\date{}

\begin{document} 


\baselineskip24pt


\maketitle


\begin{sciabstract}
  Hypervelocity stars (HVS) travel with velocities so high, that they exceed the escape velocity of the Galaxy {\it (1-3)}. Several acceleration mechanisms have been discussed. Only one HVS (US\,708, HVS\,2) is a compact helium star {\it (2)}. Here we present a spectroscopic and kinematic analysis of US\,708. Travelling with a velocity of $\sim1200\,{\rm km\,s^{-1}}$, it is the fastest unbound star in our Galaxy. In reconstructing its trajectory, the Galactic center becomes very unlikely as an origin, which is hardly consistent with the most favored ejection mechanism for the other HVS. Furthermore, we discovered US\,708 to be a fast rotator. According to our binary evolution model it was spun-up by tidal interaction in a close binary and is likely to be the ejected donor remnant of a thermonuclear supernova.
\end{sciabstract}


According to the widely-accepted theory for the acceleration of HVS, a close binary is disrupted by the supermassive black hole (SMBH) in the centre of our Galaxy and one component is ejected as hypervelocity star {\it (4)}. In an alternative scenario US\,708 was proposed to be ejected from an ultra-compact binary star by a thermonuclear supernova type Ia (SN\,Ia) {\it (5)}. However, previous observational evidence was insufficient to put firm constraints on its past evolution. Here we show that US\,708 is the fastest unbound star in our Galaxy, provide evidence for the supernova ejection scenario and identify a progenitor population of SN\,Ia.

In contrast to all other known HVS US\,708 has been classified as hot subdwarf star (sdO/B). Those stars are evolved, core helium-burning objects with low masses around $0.5\,M_{\rm \odot}$. About half of the sdB stars reside in close binaries with periods ranging from $\sim0.1\,{\rm d}$ to $\sim30\,{\rm d}$ {\it (6,7)}. The hot subdwarf is regarded as the core of a former red giant star that has been stripped off almost all of its hydrogen envelope through interaction with a close companion star {\it (8,9)}. Apparently single hot subdwarf stars like US\,708 itself are known as well. However, binary evolution has also been proposed in this case, since the merger of two helium white dwarfs (He-WDs) is a possible formation channel for those objects {\it (10)}. 

The hot subdwarf nature of US\,708 poses a particular challenge for theories, that aim at explaining the acceleration of hypervelocity stars. Within the slingshot scenario proposed by Hills, a binary consisting of two main sequence stars is disrupted by the close encounter with the SMBH in the centre of our Galaxy. While one of the components remains in a bound orbit around the black hole, the other one is ejected with high velocity {\it (4)}. This scenario not only explains the existence of the so-called S-stars orbiting the supermassive black hole in the Galactic centre and providing the most convincing evidence for the existence of this black hole {\it (11)}. It is also consistent with the main properties of the known HVS population consisting of young main sequence stars {\it (12,13)}. However, more detailed analyses of some young HVS challenge the Galactic centre origin {\it (3,14)} and most recently, a new population of old main sequence stars likely to be HVS has been discovered. Most of those objects are also unlikely to originate from the Galactic centre, but the acceleration mechanism remains unclear {\it (15)}.

In the case of the helium-rich sdO (He-sdO) US\,708 the situation is even more complicated. In contrast to all other known HVS, which are normal main sequence stars of different ages, this star is in the phase of shell helium-burning, which only lasts for a few tens of Myr. More importantly, it has been formed by close binary interaction. To accelerate a close binary star to such high velocity with the slingshot mechanism, either a binary black hole {\it (16)} or the close encounter of a hierarchical triple system, where the distant component becomes bound to the black hole and the two close components are ejected, is necessary {\it (17)}. Similar constraints apply to the dynamical ejection out of a dense cluster, which is the second main scenario discussed to explain the HVS. 

While close binarity requires quite specific modifications of the canonical HVS scenarios, it is a necessary ingredient for an alternative scenario, where US\,708 is explained as the ejected donor remnant of a thermonuclear supernova type Ia (SN\,Ia) {\it (18,19)}. Underluminous SN\,Ia have been proposed to originate from a so-called double-detonation of a white dwarf {\it (20,21)}. In this scenario a massive white dwarf (WD) is closely orbited by a low-mass helium star. Due to a tightening of the orbit the He-star will start to transfer mass to its compact companion. After a critical amount of helium is deposited on the surface of the WD through accretion, the helium is ignited causing a detonation wave that triggers the explosion of the carbon-oxygen white dwarf itself. Indeed, the ultracompact sdB+WD binary CD$-$30$^\circ$11223 has recently been identified as progenitor candidate for such a scenario and linked to the putative ejected donor remnant US\,708 {\it (5,22)}. 

We performed a detailed spectroscopic and kinematic analysis of US\,708 based on recently obtained and archival data to trace back its origin and constrain the ejection mechanism. To determine the 3D-motion of US\,708 both the radial and tangential velocity components have to be determined. We measured the radial velocity from new spectra taken with the Keck and Palomar telescopes and compared it to archival data. Atmospheric parameters and spectroscopic distance were derived from the new spectra as well (see Fig.~1). In addition to that the proper motion was determined by combining archival positions with new measurements from the PanSTARRS survey (see Fig.~S2, Table~1). 

With a Galactic restframe velocity of $1157\pm53\,{\rm km\,s^{-1}}$ we find that US\,708 is the fastest known unbound star in our Galaxy. Its current distance is $8.5\pm1.0\,{\rm kpc}$ and it is moving away from the Galactic plane into the halo. Tracing back its trajectory and assuming no further deviations we deduced that it crossed the Galactic disc $14.0\pm3.1\,{\rm Myr}$ ago. In this way an origin in the centre of our Galaxy can be excluded with high confidence (see Fig.~2), but the origin in the Galactic disc on the other hand is fully consistent with the supernova ejection scenario. In contrast to regular SN\,Ia, double-detonation SN\,Ia with hot subdwarf donors are predicted to happen in young stellar populations {\it (5)}.

\begin{table*}
\caption{Parameters of US\,708. The uncertainty of the radial velocity is the $1\sigma$ error from a $\chi^{2}$-fit, the uncertainties in the proper motion components have been propagated from the position errors by linear regression and the uncertainties in the atmospheric parameters are bootstrap errors. The uncertainties of the other, derived parameters have been propagated from the uncertainties of the input parameters.}
\begin{center}
\begin{tabular}{llll}

\hline
\noalign{\smallskip}
Visual magnitude$^{\dag}$ & $m_{\rm g}$ & [mag] & $18.668\pm0.008$ \\
Proper motion                        & $\mu_{\rm \alpha}\cos{\delta}$ & [mas/yr] & $-8.0\pm1.8$ \\
				     & $\mu_{\rm \delta}$ & [mas/yr] & $9.1\pm1.6$ \\
Radial velocity                      & $v_{\rm helio}$ & [${\rm km\,s^{-1}}$] & $917\pm7$ \\
Galactic cartesian coordinates       & $X$ & [${\rm kpc}$] & $-14.2\pm0.7$ \\
                                     & $Y$ & [${\rm kpc}$] & $0.4\pm0.1$ \\
                                     & $Z$ & [${\rm kpc}$] & $6.2\pm0.7$ \\
Galactic velocity components         & $v_{\rm X}$ & [${\rm km\,s^{-1}}$] & $-855\pm61$ \\
                                     & $v_{\rm Y}$ & [${\rm km\,s^{-1}}$] & $643\pm77$ \\
                                     & $v_{\rm Z}$ & [${\rm km\,s^{-1}}$] & $431\pm58$ \\
Galactic rest-frame velocity         & $v_{\rm grf}$ & [${\rm km\,s^{-1}}$] & $1157\pm53$ \\
\noalign{\smallskip}
\hline
\noalign{\smallskip}
Effective temperature & $T_{\rm eff}$ & [K] & $47\,200\pm400$\\
Surface gravity & $\log{g}$           & & $5.69\pm0.09$\\
Helium abundance & $\log{y}$          & & $+2.0$\\
Nitrogen abundance & $\log{\frac{N({\rm N})}{N({\rm H})}}$ & & $-2.4\pm0.2$\\
Projected rotational velocity & $v_{\rm rot}\sin{i}$ & [${\rm km\,s^{-1}}$] & $115\pm8$\\
\noalign{\smallskip}
\hline
\noalign{\smallskip}
Mass (adopted) & $M_{\rm sdB}$ & [$M_{\rm \odot}$] & $0.3$\\
Distance       & $d$  & [kpc] & $8.5\pm1.0$\\
\hline\\
\end{tabular}
\end{center}
$^{\dag}$ Taken from the Sloan Digital Sky Survey Data Server (das.sdss.org)\\
\end{table*}

Both the current Galactic restframe velocity and the reconstructed ejection velocity from the Galactic disc ($998\pm68\,{\rm km\,s^{-1}}$) are significantly higher than published before ($\sim750\,{\rm km\,s^{-1}}$, based on radial velocity alone) {\it (2)}. This puts new constraints on the possible progenitor system, which can be derived from the observed parameters of US\,708. To reach such a high ejection velocity, the progenitor binary must have been very compact and the white dwarf companion rather massive. The likely progenitor system consisted of a compact helium star with a mass of $\sim0.3\,M_{\rm \odot}$ and a massive carbon-oxygen WD ($1.0-1.2\,M_{\rm \odot}$) with an orbital period of about $10\,{\rm min}$. We calculated the mass-transfer rate in such a binary and found that the helium is accreted by the white dwarf at a rate suitable for the double-detonation scenario ($10^{-9}-10^{-8}\,M_{\rm \odot}\mathrm{yr}^{-1}$, for details see {\it (5)}). Such ultra-short period systems with compact helium stars have indeed been observed. The eclipsing He-WD+CO-WD binary SDSS\,J065133+284423 has an orbital period of only $12\,{\rm min}$ {\it (23)}. However, the mass of the CO-WD ($0.55\,M_{\rm \odot}$) is too low for a double-detonation SN\,Ia. 

The ejection from such a close binary should leave another imprint on the remnant. We know that hot subdwarfs in compact binaries have been spun up by the tidal influence of the close companion {\it (22,24,25)} to rotational velocities significantly higher than the rotational velocities of single hot subdwarfs {\it (26,27)}. An ejected remnant is predicted to have a high $v_{\rm rot}\sin{i}$ as well {\it (28)}. And indeed we measured $v_{\rm rot}\sin{i}=115\pm8\,{\rm km\,s^{-1}}$ significantly higher than expected for a single He-sdO (see Fig.~1) {\it (27)}. However, assuming an ultracompact progenitor binary, the measured $v_{\rm rot}\sin{i}$ of the He-sdO is still about a factor of four slower than expected. A significant loss of mass and angular momentum caused by the supernova impact is predicted for main sequence companions in the standard single-degenerate SN\,Ia scenario, where mass is transferred from such a companion to a WD {\it (29-31)}. While the loss of mass and angular momentum for a more compact companion like US\,708 is expected to be smaller, our results indicate that it is still substantial. This puts important observational constraints on more detailed models.

While the observed properties of US\,708 are consistent with the supernova ejection scenario, they are hardly compatible with the slingshot mechanism, because an origin of the star in the centre of the Galaxy is very unlikely (Fig.~2, see also the additional explanation in the supplementary material). However, it has to be pointed out that the supernova ejection scenario is only applicable to such compact helium stars and cannot be invoked to explain the acceleration of the other HVS.

Providing evidence that US\,708 is the likely donor remnant of a helium double-detonation SN\,Ia we not only show that the fastest unbound stars in our Galaxy are accelerated in this way. It is also an important step forward in our understanding of SN\,Ia explosions in general. Despite the fact that those bright events are used as standard candles to measure the expansion (and acceleration) of the universe, their progenitors are still unknown. Our results suggest that due to the quite high WD masses derived for the progenitor binary, the double-detonation scenario might not only be applicable to some underluminous SN\,Ia {\it (5,21)}, but might also contribute to the population of typical SN\,Ia used as cosmic yardsticks {\it (20)}.  

Depending on the pollution by SN material, the effect of the SN impact, and the subsequent stellar evolution, the surface abundances of US\,708 might be significantly affected. UV-spectroscopy is necessary to measure the metal abundances of US\,708 and put further constraints on the extreme history of this star, which witnessed a SN from a distance of less than $0.2\,R_{\rm \odot}$. 

{\bf Acknowledgements}

We thank H.~Hirsch for providing us with the LRIS spectra.\\ 
Based on observations obtained at the W.M. Keck Observatory, which is operated as a scientific partnership among the California Institute of Technology, the University of California, and the National Aeronautics and Space Administration. The Observatory was made possible by the generous financial support of the W.M. Keck Foundation. The authors wish to recognise and acknowledge the very significant cultural role and reverence that the summit of Mauna Kea has always had within the indigenous Hawaiian community. We are most fortunate to have the opportunity to conduct observations from this mountain.\\
Based on observations at the Palomar Observatory.\\
The Pan-STARRS1 Surveys (PS1) have been made possible through contributions of the Institute for Astronomy, the University of Hawaii, the Pan-STARRS Project Office, the Max-Planck Society and its participating institutes, the Max Planck Institute for Astronomy, Heidelberg and the Max Planck Institute for Extraterrestrial Physics, Garching, The Johns Hopkins University, Durham University, the University of Edinburgh, Queen's University Belfast, the Harvard-Smithsonian Center for Astrophysics, the Las Cumbres Observatory Global Telescope Network Incorporated, the National Central University of Taiwan, the Space Telescope Science Institute, the National Aeronautics and Space Administration under Grant No. NNX08AR22G issued through the Planetary Science Division of the NASA Science Mission Directorate, the National Science Foundation under Grant No. AST-1238877, the University of Maryland, and Eotvos Lorand University (ELTE).\\
Z.H. is supported by Natural Science Foundation of China (grants no. 11390374 and 11033008).\\
E.Z. and A.I. are supported by the Deutsche Forschungsgemeinschaft (DFG) through grant HE1356/45-2.\\
T.K. acknowledges support by the Netherlands Research School for Astronomy (NOVA).\\
A.I. acknowledges support from a research scholarship by the Elite Network of Bavaria.\\
The data observed with the SDSS and Keck telescope are published via the SDSS and Keck data archive, the PS1 data and catalogue will become public at the end of 2014.\\

\newpage

{\bf Figures}

\includegraphics[width=14cm]{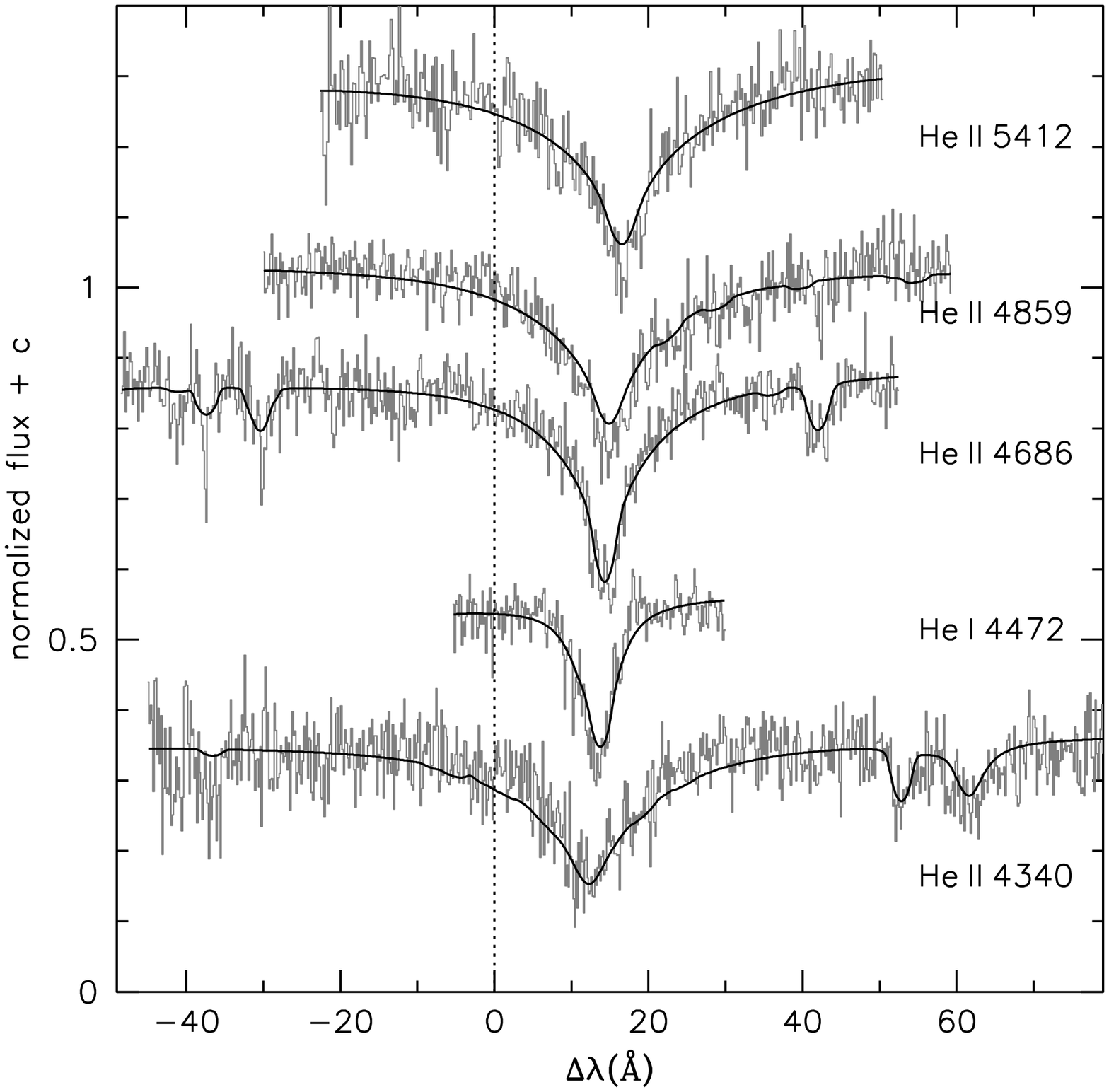}

Fig.~1: Fit of model spectrum. Fit of synthetic NLTE models to the helium and nitrogen lines of a Keck/ESI spectrum of US\,708. The normalized fluxes of the single lines are shifted for better visualisation and the most prominent lines are labeled. The weaker lines are from N\,{\sc iii} at $4634\,{\rm \AA}$, $4640\,{\rm \AA}$, He\,{\sc i} at $4713\,{\rm \AA}$ (middle plot), N\,{\sc iii} at $4379\,{\rm \AA}$ and He\,{\sc i} at $4387\,{\rm \AA}$ (bottom plot). The dashed vertical line marks the rest wavelengths of the lines. The high radial velocity shift as well as the significant broadening of the lines are clearly visible.

\includegraphics[width=14cm]{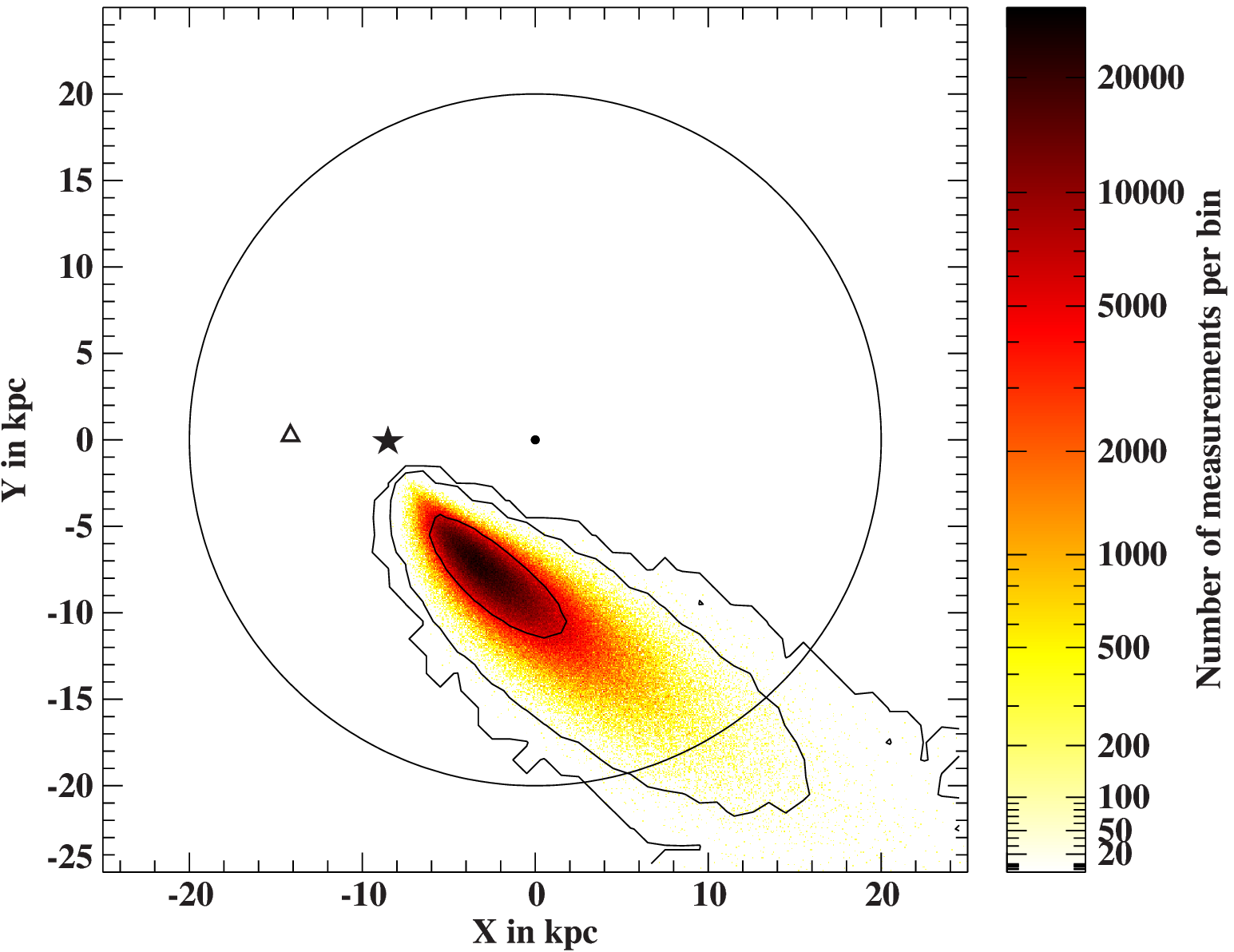}

Fig.~2: Origin of US\,708. Monte Carlo simulation ($10^{8}$ iterations) of the past trajectory of US\,708. The colour-coded bins mark the positions, where the star crossed the Galactic disc, which is shown pole-on. The contours correspond to the $1\sigma$, $3\sigma$ and $5\sigma$ confidence limits. The position of the Galactic centre is marked by the black dot, the position of the Sun with the star symbol. The current position of US\,708 is marked by a triangle and given in Table~1. 

{\bf References and Notes}

\begin{enumerate}
\item{Brown, W. R., Geller, M. J., Kenyon, S. J., \& Kurtz M. J. Discovery of an Unbound Hypervelocity Star in the Milky Way Halo. {\em Astrophys. J.} {\bf 622}, L33  (2005).}
\item{Hirsch, H. A., Heber, U., O'Toole, S. J., \& Bresolin, F. US 708 - an unbound hyper-velocity subluminous O star. {\em Astron. Astrophys.} {\bf 444}, L61 (2005).}
\item{Edelmann, H., Napiwotzki, R., Heber, U., Christlieb, N. \& Reimers, D. HE 0437-5439: An Unbound Hypervelocity Main-Sequence B-Type Star {\em Astrophys. J.} {\bf 634}, L181 (2005).}
\item{Hills, J. G. Hyper-velocity and tidal stars from binaries disrupted by a massive Galactic black hole. {\em Nature} {\bf 331}, 687 (1988).}
\item{Geier, S., et al. A progenitor binary and an ejected mass donor remnant of faint type Ia supernovae. {\em Astron. Astrophys.} {\bf 554}, 54 (2013).}
\item{Maxted, P. F. L., Heber, U., Marsh, T. R., \& North, R. C. The binary fraction of extreme horizontal branch stars. {\em Mon. Not. R. Astron. Soc.} {\bf 326}, 1391 (2001).}
\item{Napiwotzki, R., et al. Close binary EHB stars from SPY. {\em Astrophysics and Space Science} {\bf 291}, 321 (2004).}
\item{Han, Z., Podsiadlowski, Ph., Maxted, P. F. L., Marsh, T. R., \& Ivanova, N. The origin of subdwarf B stars -- I. The formation channels. {\em Mon. Not. R. Astron. Soc.} {\bf 336}, 449 (2002).}
\item{Han, Z., Podsiadlowski, Ph., Maxted, P. F. L., \& Marsh, T. R. The origin of subdwarf B stars -- II. {\em Mon. Not. R. Astron. Soc.} {\bf 341}, 669 (2003).}
\item{Webbink, R. F. Double white dwarfs as progenitors of R Coronae Borealis stars and Type I supernovae. {\em Astrophys. J.} {\bf 279}, 252 (1984).}
\item{Gillessen, S., et al. Monitoring Stellar Orbits Around the Massive Black Hole in the Galactic Center. {\em Astrophys. J.} {\bf 692}, 1075 (2009).}
\item{Brown, W. R., Cohen, J. G., Geller, M. J., \& Kenyon, S. J. The Nature of Hypervelocity Stars and the Time between Their Formation and Ejection. {\em Astrophys. J.} {\bf 754}, L2  (2012).}
\item{Bromley, B. C., Kenyon, S. J., Geller, M. J., \& Brown, W. R. Binary Disruption by Massive Black Holes: Hypervelocity Stars, S Stars, and Tidal Disruption Events. {\em Astrophys. J.} {\bf 749}, L42 (2012).}
\item{Heber, U., Edelmann, H., Napiwotzki, R., Altmann, M., \& Scholz, R.-D. The B-type giant HD 271791 in the Galactic halo. Linking run-away stars to hyper-velocity stars. {\em Astron. Astrophys.} {\bf 483}, L21 (2008).}
\item{Palladino, L. E., et al. Hypervelocity star candidates in the SEGUE G and K dwarf sample. {\em Astrophys. J.} {\bf 780}, 7 (2014).}
\item{Lu, Y., Yu, Q., \& Lin, D. N. C. Hypervelocity Binary Stars: Smoking Gun of Massive Binary Black Holes. {\em Astrophys. J.} {\bf 666}, L89 (2007).}
\item{Perets, H. P. Runaway and Hypervelocity Stars in the Galactic Halo: Binary Rejuvenation and Triple Disruption. {\em Astrophys. J.} {\bf 698}, 1330 (2009).}
\item{Wang, B., Meng, X., Chen, X., \& Han, Z. The helium star donor channel for the progenitors of Type Ia supernovae. {\em Mon. Not. R. Astron. Soc.} {\bf 395}, 847 (2009).}
\item{Justham, S., Wolf, C., Podsiadlowski, Ph., \& Han, Z. Type Ia supernovae and the formation of single low-mass white dwarfs. {\em Astron. Astrophys.} {\bf 493}, 1081 (2009).}
\item{Fink, M., et al. Double-detonation sub-Chandrasekhar supernovae: can minimum helium shell masses detonate the core? {\em Astron. Astrophys.} {\bf 514}, 53 (2010).}
\item{Foley, R., et al. Type Iax Supernovae: A New Class of Stellar Explosion. {\em Astrophys. J.} {\bf 767}, 57 (2013).}
\item{Vennes, S., Kawka, A., O'Toole, S. J., N\'emeth, P, \& Burton, D. The Shortest Period sdB Plus White Dwarf Binary CD-30 11223 (GALEX J1411-3053). {\em Astrophys. J.} {\bf 759}, L25  (2012).}
\item{Brown, W. R., et al. A 12 Minute Orbital Period Detached White Dwarf Eclipsing Binary. {\em Astrophys. J.} {\bf 737}, L23 (2011).}
\item{Geier, S., et al. The hot subdwarf B + white dwarf binary KPD\,1930$+$2752 -- A supernovae type Ia progenitor candidate. {\em Astron. Astrophys.} {\bf 464}, 299 (2007).}
\item{Geier, S., et al. Hot subdwarf stars in close-up view. I. Rotational properties of subdwarf B stars in close binary systems and nature of their unseen companions. {\em Astron. Astrophys.} {\bf 519}, 25 (2010).}
\item{Geier, S., \& Heber, U. Hot subdwarf stars in close-up view. II. Rotational properties of single and wide binary subdwarf B stars. {\em Astron. Astrophys.} {\bf 543}, 149 (2012).}
\item{Heber, U., \& Hirsch, H. A. Carbon abundances of sdO stars from SPY. {\em Journal of Physics Conf. Ser.} {\bf 172}, 012015 (2009).}
\item{Pan, K.-C., Ricker, P. M., \& Taam, R. E. Evolution of Post-impact Remnant Helium Stars in Type Ia Supernova Remnants within the Single-degenerate Scenario. {\em Astrophys. J.} {\bf 773}, 49 (2013).}
\item{Pan, K.-C., Ricker, P. M., \& Taam, R. E. Impact of Type Ia Supernova Ejecta on Binary Companions in the Single-degenerate Scenario. {\em Astrophys. J.} {\bf 750}, 151 (2012).}
\item{Pan, K.-C., Ricker, P. M., \& Taam, R. E. Evolution of Post-impact Companion Stars in SN Ia Remnants within the Single-degenerate Scenario. {\em Astrophys. J.} {\bf 760}, 21 (2012).}
\item{Liu, Z.-W., et al. Rotation of surviving companion stars after type Ia supernova explosions in the WD+MS scenario. {\em Astron. Astrophys.} {\bf 554}, 109 (2013).}
\item{Ahn, C. P., et al. The Ninth Data Release of the Sloan Digital Sky Survey: First Spectroscopic Data from the SDSS-III Baryon Oscillation Spectroscopic Survey. {\em Astrophys. J. Suppl.} {\bf 203}, 21 (2012).}
\item{Napiwotzki, R., et al. Double degenerates and progenitors of supernovae type Ia. {\em ASP Conf. Series} {\bf 318}, 402 (2004).}
\item{Tillich, A., et al. The Hyper-MUCHFUSS project: probing the Galactic halo with sdB stars. {\em Astron. Astrophys.} {\bf 527}, 137 (2011).}
\item{Geier, S., et al. The MUCHFUSS project - searching for hot subdwarf binaries with massive unseen companions. Survey, target selection and atmospheric parameters. {\em Astron. Astrophys.} {\bf 530}, 28 (2011).}
\item{N\'emeth, P., Kawka, A., \& Vennes, S. A selection of hot subluminous stars in the GALEX survey - II. Subdwarf atmospheric parameters. {\em Mon. Not. R. Astron. Soc.} {\bf 427}, 2180  (2012).}
\item{Ahmad, A., \& Jeffery, C. S. Physical parameters of helium-rich subdwarf B stars from medium resolution optical spectroscopy. {\em Astron. Astrophys.} {\bf 402}, 335 (2003).}
\item{Naslim, N., Jeffery, C. S., Ahmad, A., Behara, N. T. \& Sahin, T. Abundance analyses of helium-rich subluminous B stars. {\em Mon. Not. R. Astron. Soc.} {\bf 409}, 582 (2010).}
\item{Stroeer, A., et al. Hot subdwarfs from the ESO supernova Ia progenitor survey. II. Atmospheric parameters of subdwarf O stars. {\em Astron. Astrophys.} {\bf 462}, 269 (2007).}
\item{Ramspeck, M., Heber, U., \& Moehler, S. Early type stars at high galactic latitudes. I. Ten young massive B-type stars. {\em Astron. Astrophys.} {\bf 378}, 907 (2001).}
\item{Schlegel, D. J., Finkbeiner, D. P., \& Davis, M. Maps of Dust Infrared Emission for Use in Estimation of Reddening and Cosmic Microwave Background Radiation Foregrounds. {\em Astrophys. J.} {\bf 500}, 525 (1998).}
\item{Allen, C. \& Santillan, A. An improved model of the galactic mass distribution for orbit computations. {\em Revista Mexicana de Astronomia y Astrofisica} {\bf 22}, 255 (1991).}
\item{Irrgang, A., Wilcox, B., Tucker, E., \& Schiefelbein, L. Milky way mass models for orbit calculations. {\em Astron. Astrophys.} {\bf 549}, 137 (2013).}
\item{Eggleton, P. Approximations to the radii of Roche lobes. {\em Astrophys. J.} {\bf 268}, 363 (1983).}
\item{Han, Z., \& Webbink. R. F. Stability and energetics of mass transfer in double white dwarfs. {\em Astron. Astrophys.} {\bf 349}, L17 (1999).}
\item{Brown, W. R., Kilic, M., Allende Prieto, C., Gianninas, A., \& Kenyon, S. J. The ELM Survey. V. Merging Massive White Dwarf Binaries. {\em Astrophys. J.} {\bf 769}, 66 (2013).}
\item{Driebe, T., Sch\"onberner, D., Bl\"ocker, T., \& Herwig, F. The evolution of helium white dwarfs. I. The companion of the millisecond pulsar PSR J1012+5307. {\em Astron. Astrophys.} {\bf 339}, 123 (1998).}
\item{Marsh, T. R., et al. Detection of a White Dwarf Companion to the White Dwarf\\ SDSSJ\,125733.63+542850.5. {\em Astrophys. J.} {\bf 736}, 95 (2011).}
\item{Geier, S., Classen, L., \& Heber, U. The Fast-rotating, Low-gravity Subdwarf B Star EC 22081-1916: Remnant of a Common Envelope Merger Event. {\em Astrophys. J.} {\bf 733}, L13 (2011).}
\item{Geier, S., et al. The subdwarf B star SB 290 - A fast rotator on the extreme horizontal branch. {\em Astron. Astrophys.} {\bf 551}, L4 (2013).}
\item{Liu, Z.-W., Pakmor, R., Seitenzahl, I. R., et al. The Impact of Type Ia Supernova Explosions on Helium Companions in the Chandrasekhar-mass Explosion Scenario. {\em Astrophys. J.} {\bf 774}, 37 (2013).}
\item{Blaauw, A. On the origin of the O- and B-type stars with high velocities (the "run-away" stars), and some related problems. {\em Bull. Astron. Inst. Neth.} {\bf 15}, 265 (1961).}
\item{Gvaramadze, V. V., Gualandris, A., \& Protegies Zwart, S. Hyperfast pulsars as the remnants of massive stars ejected from young star clusters. {\em Mon. Not. R. Astron. Soc.} {\bf 385}, 929 (2008).}
\item{Abadi, M. G., Navarro, J. F., \& Steinmetz, M. An Alternative Origin for Hypervelocity Stars. {\em Astrophys. J.} {\bf 691}, 63 (2009).}
\item{Geier, S., et al. The MUCHFUSS Project: Searching for the Most Massive Companions to Hot Subdwarf Stars in Close Binaries and Finding the Least Massive Ones. {\em ASP Conf. Ser.} { \bf 452}, 129 (2012).}
\item{Wang, B., et al. Birthrates and delay times of Type Ia supernovae. {\em Sci. China Ser. G} {\bf 53}, 586 (2010).} 
\item{Yu, G., \& Tremaine, S. Ejection of Hypervelocity Stars by the (Binary) Black Hole in the Galactic Center. {\em Astrophys. J.} {\bf 599}, 1129 (2003).}
\end{enumerate}

\newpage
\begin{center}

{\Huge Supplementary Materials for}\\
\vspace{1cm}
{\Large The fastest unbound star in our Galaxy ejected by a thermonuclear supernova}\\
\vspace{0.5cm}
S. Geier, F. F\"urst, E. Ziegerer, T. Kupfer, U. Heber,\\
A. Irrgang, B. Wang, Z. Liu, Z. Han, B. Sesar, D. Levitan,\\
R. Kotak, E. Magnier, K. Smith, W. S. Burgett,\\ 
K. Chambers, H. Flewelling, N. Kaiser, R. Wainscoat, \\
C. Waters\\
\vspace{0.5cm}
correspondence to: sgeier@eso.org\\
\vspace{1.0cm}

\end{center}

{\Large This PDF file includes:}

{\large Materials and Methods}

{\large Supplementary Text}

{\large Figs. S1 to S7}

\section*{Materials and Methods}

\subsection*{Summary}

Spectra of US\,708 have been taken with the 10m Keck and the 5m Palomar telescopes. From the Doppler shift of the spectral lines we measured the radial velocity using both new and archival data. The tangential velocity components have been determined by measuring the proper motion of the star from multi-epoch position measurements spanning 59 years and its spectroscopic distance performing a full quantitative spectral analysis using state-of-the-art model atmospheres. Using those informations we constrained the kinematics of this star and traced back its origin to the Galactic disk performing a Monte Carlo simulation. The properties of the progenitor binary have been determined mostly based on the derived ejection velocity from the Galactic disc. Binary evolution calculations have then been performed to check the consistency of those properties with theory. The current rotational properties of US\,708 have been compared with hydrodynamical models of angular momentum-loss triggered by supernovae explosions. 

\subsection*{Observations} 

US\,708 ($\alpha_{2000}=09^{\rm h}33^{\rm m}20{\stackrel{\rm s}{\displaystyle.}}85$, $\delta_{2000}=+44^{\rm \circ}17'05.8''$) was discovered to be a hypervelocity star by Hirsch et al. {\it (2)}. A medium-resolution ($R\sim1800$) spectrum was taken by the Sloan Digital Sky Survey (SDSS) on February 20, 2002 {\it (32)}. Follow-up low-resolution ($R\sim900$) spectroscopy was obtained with the LRIS instrument at the Keck telescope on May 13, 2005. The reduced spectra from the blue and red channel of the instrument were provided to us by H. Hirsch. A series of 11 consecutively taken medium-resolution ($R\sim8000$) spectra was obtained with the ESI instrument at the Keck telescope on March 3, 2013. The spectra have been reduced with the ESI pipeline Makee.\footnote{http://www.astro.caltech.edu/ \textasciitilde tb/makee/} One spectrum has been taken with the medium-resolution spectrograph at the 5.1\,m Hale telescope on Mount Palomar on May 11, 2013 and another three spectra on June 1, 2013.

\subsection*{Revised radial velocity} 

Hirsch et al. measured the radial velocity (RV) of US\,708 ($708\pm15\,{\rm km\,s^{-1}}$) from the helium lines in the blue-channel LRIS spectrum. { The measured RV exceeded the typical RVs of He-sdOs in the rest of the sample, which is characteristic for halo stars, considerably (see Fig.~5 in {\it (35)}).} We obtained the RV of the SDSS, ESI and Palomar spectra by fitting a model spectrum (see below) to the helium lines using the FITSB2 routine {\it (33)}. Surprisingly, the most accurate RV measured from the coadded ESI spectrum ($917\pm7\,{\rm km\,s^{-1}}$) turned out to be significantly higher than the one published by Hirsch et al. {\it (2)}. This velocity is consistent with the RVs measured both from the SDSS\footnote{The RV of $793\,{\rm km\,s^{-1}}$ given by the SDSS Sky Server Object Explorer tool is based on the fit of an hydrogen rich template to the spectrum. The He\,{\sc ii}-lines of the Pickering series are misidentified as Balmer lines. This introduces the shift of $\sim-100\,{\rm km\,s^{-1}}$ between our result and the template fit.} ($898\pm30\,{\rm km\,s^{-1}}$) and the Palomar spectra ($866-936\,{\rm km\,s^{-1}}$). To investigate this issue, we performed a reanalysis of the LRIS spectra and measured an RV of $709\pm7\,{\rm km\,s^{-1}}$ for the LRIS blue-channel spectrum perfectly consistent with the published value. However, when fitting the red-channel spectrum we found a significantly discrepant RV of $797\pm21\,{\rm km\,s^{-1}}$. This was taken as first indication, that those spectra might be affected by systematics. We used the nightsky emission line of O\,{\sc i} at $6300\,{\rm \AA}$ (red-channel) and the interstellar absorption line of Ca\,{\sc ii} at $3934\,{\rm \AA}$ (blue-channel) to quantify those systematic shifts (see Fig.~S1, left panel). The nightsky emission line, which is supposed to be at zero RV, was blue-shifted by $-33\pm10\,{\rm km\,s^{-1}}$. The interstellar line showed a significantly higher shift of $-128\pm22\,{\rm km\,s^{-1}}$. Since the correct RV of the interstellar line is not known a priori, we measured it from a coadded Palomar spectrum to be $-9\pm28\,{\rm km\,s^{-1}}$. Correcting the RVs for those shifts, the two RV values ($828\pm22\,{\rm km\,s^{-1}}$ blue-channel, $830\pm21\,{\rm km\,s^{-1}}$ red-channel) from the LRIS spectra are consistent with each other, but still smaller than the RVs measured from the ESI and Palomar spectra (see Fig.~S1, right panel). Due to the low resolution of the LRIS spectra, the remaining shift corresponds to only about one pixel on the CCD and is therefore regarded as systematic as well. We conclude that the RV published by Hirsch et al. was affected by systematics and therefore underestimated. Going back to the original raw data, we performed an independent reduction. However, we were not able to resolve this issue. Given that the uncertainties are at the $1\sigma$ level of confidence and that the LRIS spectra are affected by systematics, no significant shifts in RV on both short and long timescales are detected (see Supplementary Fig.~1, right panel). 

\subsection*{Proper motion} 

The proper motion of US\,708 has been derived from multi-epoch position measurements of Schmidt plates obtained from the Digitised Sky Survey (DSS)\footnote{http://archive.stsci.edu/cgi-bin/dss\_plate\_finder}, the Sloan Digital Sky Survey {\it (32)} and the PanSTARRS survey (PS1) over a timebase of 59 years (see Fig.~S2). The positions from the DSS and SDSS images have been measured with respect to a set of compact background galaxies selected from the SDSS images as described in Tillich et al. {\it (34)}. The positions of the background galaxies and the object are measured. For each image the measured positions of the background galaxies are compared to the reference values from the PS1 catalogue. The average of the deviations from these reference values is adopted as uncertainty of the object position. In the case of the 29 PS1 epochs, we took the calibrated positions from the PS1 catalogue and therefore use the same reference system for all our measurements. We obtained one position per epoch and used linear regression ro derive the proper motion components with their uncertainties.

\subsection*{Atmospheric parameters} 

The atmospheric parameters effective temperature $T_{\rm eff}$, surface gravity $\log\,g$, nitrogen abundance $\log\,N({\rm N})/N({\rm H})$ and projected rotational velocity $v_{\rm rot}\sin{i}$ were determined by fitting simultaneously the observed helium and nitrogen lines of an ESI spectrum, constructed by coadding the 11 single exposures, with NLTE models taking into account line-blanketing of nitrogen {\it (27)} (see Fig.~1) as described in Geier et al. {\it (35)}. Since no hydrogen lines are visible in the spectrum, we fixed the helium abundance to $\log\,y=\log\,N({\rm He})/N({\rm H})=+2.0$. The atmospheric parameters ($T_{\rm eff}=47200\pm400\,{\rm K}$, $\log{g}=5.69\pm0.09$) deviate significantly from the results by Hirsch et al. ($T_{\rm eff}=45600\pm700\,{\rm K}$, $\log{g}=5.23\pm0.12$) especially in surface gravity. This is caused by the additional line-blanketing of nitrogen {\it (27)}. The atmospheric parameters as well as the nitrogen abundance $\log\,N({\rm N})/N({\rm H})=-2.4\pm0.2$ are typical for the nitrogen-rich subclass of He-sdOs {\it (36)}. The projected rotational velocity $v_{\rm rot}\sin{i}=115\pm8\,{\rm km\,s^{-1}}$ on the other hand is significantly higher than the ones of both single sdB ($<10\,{\rm km\,s^{-1}}$) and He-sdO stars ($20-30\,{\rm km\,s^{-1}}$) {\it (26,27)}. { Based on our analysis we can rule out objects with similar spectral features like DO-type white dwarfs ($\log{g}>7.0$) or luminous He-stars ($\log{g}<4.5$), which can be easily misclassified from visual inspection only.}

\subsection*{Spectroscopic distance and kinematics} 

The spectroscopic distance is derived from the atmospheric parameters $T_{\rm eff}$, $\log\,g$ and the apparent visual magnitude as described in Ramspeck et al. {\it (40)}. The SDSS-g and r magnitudes have been converted to Johnson V magnitude\footnote{http://www.sdss.org/dr6/algorithms/sdssUBVRITransform.html}, which has been  corrected for interstellar reddening ($A_{\rm V}=0.07\,{\rm mag}$) {\it (41)}. Based on constraints provided by the supernova ejection scenario (see below) we adopted a mass of $0.3\,M_{\rm \odot}$ for the hot subdwarf. The spectroscopic distance in this case is $8.5\pm1.0\,{\rm kpc}$. For this distance the proper motion components are converted to absolute transverse velocities and, combined with the radial velocity, the Galactic restframe velocity of US\,708 is calculated to be $1157\pm53\,{\rm km\,s^{-1}}$. This is the highest known restframe velocity of any unbound star in our Galaxy. The past trajectory of US\,708 in the Galactic potential {\it (42,43)} has been reconstructed as outlined in Tillich et al. {\it (34)}. Due to the high velocity of US\,708 we found that the trajectory is not changed significantly if alternative model potentials are used {\it (42)}. US\,708 was ejected from the Galactic disc $14.0\pm3.1\,{\rm Myr}$ ago and the ejection velocity, corrected for the motion of the Galactic disc, was calculated to be $998\pm68\,{\rm km\,s^{-1}}$. We performed Monte Carlo simulations to trace back the trajectory of US\,708 until it intersects with the Galactic disc. The uncertainty in the proper motion measurement dominates the error budget. Assuming no further perturbation of the trajectory an origin from the central kpc around the Galactic centre can be ruled out with a confidence of more than $6\sigma$ (see Fig.~2). 

{ Properties of the progenitor binary.} Geier et al. {\it (5)} proposed the ultra-compact sdB+WD binary CD$-$30$^\circ$11223 to be a possible progenitor of the hypervelocity sdO US\,708. However, based on the new results presented here, the ejection velocity is $\sim250\,{\rm km\,s^{-1}}$ higher than assumed by Geier et al. Hence, it is necessary to reexamine the supernova ejection scenario and to test its consistency with the newly derived parameters of US\,708. Similar to the scenario discussed in Geier et al., we assume that the progenitor binary consisted of a compact helium star and a massive carbon-oxygen WD in close orbit. The ejection velocity of the He-star equals the radial velocity semiamplitude of the progenitor binary at the moment of the supernova explosion ($K=998\pm68\,{\rm km\,s^{-1}}$) modified by the additional perpendicular velocity component the star received through the SN explosion ($\sim200\,{\rm km\,s^{-1}}$). Since both velocities are added in quadrature, the kick velocity is negligible within the uncertainties. Assuming a mass for the He-star and a circular orbit, the orbital period of the progenitor binary as well as its separation can be calculated from the binary mass function:

\begin{equation}
 \label{equation-mass-function}
 f_{\rm m} = \frac{M_{\rm WD}^3 \sin^3i}{(M_{\rm WD} +
   M_{\rm He})^2} = \frac{P K^3}{2 \pi G}
\end{equation}

Since we known the absolute space velocity, the inclination angle can be set to $\sin{i}=1$, and the orbital period $P$ of the progenitor binary can be calculated:

\begin{equation}
 \label{equation-period}
 P = \frac{2 \pi G}{K^3}\frac{M_{\rm WD}^3}{(M_{\rm WD} +
   M_{\rm He})^2}
\end{equation}

The binary separation $a$ can be derived using Keplers laws:

\begin{equation}
 \label{equation-separation}
 a = \frac{G}{K^2}\frac{M_{\rm WD}^2}{M_{\rm WD} +
   M_{\rm He}}
\end{equation}

Another crucial assumption is that stable mass-transfer from the He-star to the WD triggered the SN, which means that the He-star must have filled its Roche lobe before ejection. To calculate the Roche lobe radius we used the equation given by Eggleton {\it (44)}, where $q=M_{\rm He}/M_{\rm M_{\rm WD}}$:

\begin{equation}
 \label{equation-roche}
  R_{\rm L} = \frac{0.49q^{2/3}a}{0.6q^{2/3}+\ln(1+q^{1/2})}
\end{equation}

The radius of the He-star can be calculated as a function of the mass and the surface gravity $g$:

\begin{equation}
R_{\rm He}=\sqrt{\frac{M_{\rm He}G}{g}}
\end{equation}
 
To compare the Roche radius with the possible radius of the He-star we have to take into account that US\,708 has already evolved away from the EHB (see Fig.~S3), which led to an increase in radius. To calculate the radius at the time of ejection we therefore adopt the highest reasonable values for $\log{g}\simeq6.1$ close to the ZAEHB and the He-MS. Calculating Roche lobe and He-star radii for different He-star and WD masses, we explored the parameter space and put constraints on possible progenitor systems. Fig.~S4 shows the Roche radii for WD masses from $1.0\,M_{\rm \odot}$ to $1.2\,M_{\rm \odot}$. The He-star radii for $\log{g}=6.1$ are plotted for comparison. Consistent solutions are only found for low He-star masses ($\sim0.3-0.35\,M_{\rm \odot}$) and high WD companion masses ($\sim1.0-1.2\,M_{\rm \odot}$). The orbital period of the progenitor binary can be constrained to $\sim10\,{\rm min}$. Following the method described in Geier et al. {\it (5)} we calculated the mass-transfer rates for binaries with similar orbital parameters and component masses. The rates of $\sim10^{-8}\,M_{\rm \odot}\mathrm{yr}^{-1}$ are consistent with the helium double-detonation scenario. Figs.~S5-S7 show as an example the evolution of a close binary ($P=26\,{\rm min}$) that starts mass-transfer with an initial He-star mass of $0.45\,M_{\rm \odot}$ and a WD mass of $1.05\,M_{\rm \odot}$. After about $5\,{\rm Myr}$ the orbital period becomes as short as $13\,{\rm min}$ and the component masses change to $0.3\,M_{\rm \odot}$ and $1.2\,M_{\rm \odot}$, when the WD explodes as SN\,Ia. The helium donor has to be a helium-burning star rather than a He-WD without ongoing nuclear burning in its core, because those objects consist of degenerate matter and as soon as the small non-degenerate envelope is transferred, the mass-transfer rate becomes too high for the double-detonation scenario. Such systems will experience He-flashes on the surface of the WD companion without igniting the carbon in the core {\it (45)}. This is consistent with the observational evidence. Since the ejection already happened $\sim14\,{\rm Myr}$ ago, we can also assume that US\,708 is a helium-burning star. The minimum mass for such objects is $\sim0.3\,M_{\rm \odot}$. Even less massive He-stars without nuclear burning, which are the progenitors of He-WDs, exist {\it (46)}. However, according to evolutionary tracks, their effective temperatures are much lower than the one of US\,708. The most massive He-WD progenitors on the other hand, which can reach such temperatures, are cooling within a few Myr, too fast to be consistent with the position of US\,708 in the $T_{\rm eff}-\log{g}$-diagram (see Fig.~S3) {\it (47)}. EHB-tracks for masses as low as $\sim0.3\,M_{\rm \odot}$ are also not consistent with the position of US\,708 in the $T_{\rm eff}-\log{g}$-diagram {\it (6)}. However, since the He-star was significantly more massive before the mass-transfer started, its further evolution might not depend on its current total mass. Especially, if the helium in its core was already exhausted towards the end of the mass-transfer phase, the further evolution would depend on the core mass rather than the total mass. The position of US\,708 in the $T_{\rm eff}-\log{g}$-diagram is, for example, perfectly consistent with post-EHB model tracks for an original mass of $0.45\,M_{\rm \odot}$ (see Fig.~S3). Based on these simple calculations we can rule out the sdB+WD binary CD$-$30$^\circ$11223 ($0.51\,M_{\rm \odot} + 0.76\,M_{\rm \odot}$, $P=72\,{\rm min}$) as direct progenitor of US\,708. However, systems like CD$-$30$^\circ$11223 remain progenitor candidates of other high velocity sdB stars {\it (34)}. There is no binary known yet, which fulfills all the criteria for a progenitor system to US\,708. But evidence is growing, that such objects exist. A whole population of close binaries consisting of He-WDs and CO-WDs has been discovered recently {\it (46)}. They form in almost exactly the same way as sdB+WD binaries. The only difference is that core helium-burning was not ignited before the hydrogen envelope has been stripped off in the common envelope phase of unstable mass-transfer. The ultracompact, eclipsing He-WD+CO-WD ($0.25\,M_{\rm \odot}+0.55\,M_{\rm \odot}$) binary SDSS\,J065133+284423 with an orbital period of only $12\,{\rm min}$ sticks out {\it (23)}. Its period is similar to the one expected for the progenitor of US\,708, but the component masses are too small. In the double-lined WD+WD binary SDSS\,J125733+542850 ($0.2\,M_{\rm \odot}+1.2\,M_{\rm \odot}$) on the other hand, the masses are very similar to the ones predicted, whereas the orbital period is much longer ($274\,{\rm min}$) {\it (48)}. These discoveries as well as binary evolution calculations indicate the existence of binaries fulfilling the criteria for a progenitor of US\,708 as well {\it (18)}. We therefore conclude that the double-detonation supernova ejection scenario is still able to explain the observed properties of US\,708 as ejected donor remnant. 

\subsection*{Rotational velocity} 

Only two out of more than $100$ single hot subdwarf stars are fast rotators. Both objects are sdB stars with hydrogen-rich atmospheres and might have been formed by a common-envelope merger {\it (49,50)}. US\,708 on the other hand belongs to the population of He-sdOs, which are hotter and show no or only some hydrogen in their atmospheres. They are regarded as a distinct group of stars, that might have been formed in different ways as the sdBs. US\,708 is the only single He-sdO rotating faster than $20-30\,{\rm km\,s^{-1}}$ indicating a close-binary origin {\it (27)}. Due to the short orbital period and high companion mass, the rotation of the He-star in the proposed progenitor binary is expected to be synchronised to its orbital motion {\it (22,24,25)}. Assuming the angular momentum is unchanged after the SN, the ejected donor remnant should remain a fast rotator. The rotational velocity can be calculated as follows:

\begin{equation}
v_{\rm rot}=\frac{2\pi R_{\rm L}}{P}
\end{equation}

The expected rotational velocity of the ejected He-star is only weakly dependent on the masses of the binary components and of the order of $600\,{\rm km\,s^{-1}}$. This is much higher than the measured $v_{\rm rot}\sin{i}=115\pm8\,{\rm km\,s^{-1}}$. The significant difference between the expected rotational velocity and the measured $v_{\rm rot}\sin{i}$ comes unexpected. In a synchronised binary system the rotational axes of both components are perpendicular to the orbital plane. As soon as the He-star is ejected, the rotation axis should be perpendicular to the flight trajectory, which means that $\sin{i}\simeq1$. The impact of the supernova shockwave on main sequence (MS) companions in the standard single-degenerate scenario has recently been studied with hydrodynamic simulations. Due to stripping of matter, the star may lose up to $\sim90\%$ of its initial angular momentum. A subsequent increase in radius due to stellar evolution is also predicted to lower the rotational velocity at the surface {\it (28-31)}. However, simulations of more compact helium stars show that much less mass is stripped {\it (28,51)}. The loss of angular momentum is also expected to be smaller in this case. Taking into account evolution on the extreme horizontal branch (EHB), the radius of the sdO increased by about a factor of $\sim1.6$ since the ejection. Assuming conservation of angular momentum, the rotational velocity should now be of the order of $400\,{\rm km\,s^{-1}}$. Whether the rest of the angular momentum was lost in the SN impact or later is still unclear. Pan et al. {\it (28)} predict an increase of the helium star's radius by a factor of up to four right after the impact. This phase should last for a few tens of years. Due to the high initial rotational velocity of the star, another episode of mass and angular momentum loss may be possible in this phase. Another possibility might be a tilting of the stars rotation axis by the SN impact. The projected rotational velocity of the star measured from the line broadening would then be smaller than the true rotational velocity. However, simulations show that this effect is negligible for MS stars and most likely also for the more compact He stars studied here.

\section*{Supplementary Text}

\subsection*{Discussing alternative acceleration scenarios}

Any scenario for the acceleration of US\,708 must explain four key properties of this star simultaneously: (I) US\,708 is a compact He-sdO, which most likely formed via close binary interaction. Either it is the stripped core of a red giant or the result of a He-WD merger. (II) The star has the highest Galactic restframe velocity ($\sim1200\,{\rm km\,s^{-1}}$) ever measured for any unbound star in the Galaxy. (III) The past trajectory of the star is not consistent with an origin in the Galactic centre. (IV) In contrast to all other known single He-sdOs, US\,708 has a projected rotational velocity exceeding $\sim100\,{\rm km\,s^{-1}}$. We now want to review other scenarios for the acceleration of hypervelocity stars in this way.
In the classical runaway scenario a massive star in a binary system explodes as core-collapse supernova, while the companion is kicked out of the system {\it (52)}. However, the ejection velocity scales with the binary separation and because a massive and hence large star is involved, the binary separation cannot become small enough to reach an ejection velocity like the one of US\,708. The disruption of a hierarchical triple system consisting of a normal star in wide orbit around a close He-WD binary by the SMBH { is regarded as very unlikely} as well, because an origin in the GC { where the SMBH is located} is very unlikely. The subsequent merger of a He-WD binary ejected in this way has been proposed as formation scenario for US\,708 {\it (2,16)}. The ejection of a He-WD binary star by a hypothetical binary black hole { in the GC is very unlikely for the same reason} {\it (16)}. Other formation channels for hypervelocity stars invoke dynamical interactions in dense stellar clusters {\it (53)}. Interactions of two close binaries can lead to the ejection of a star with the appropriate velocity. However, the binary-binary interaction is not affecting the angular momentum of the ejected star. We can therefore assume that the observed $v_{\rm rot}\sin{i}=115\pm8\,{\rm km\,s^{-1}}$ resembles the rotational velocity in the tidally-locked progenitor binary. To reach such a high rotational velocity, the separation of this binary is constrained to $\sim1\,R_{\rm \odot}$ {\it (24)}. The interaction probability of two such binaries even in a very dense cluster is expected to be extremely small. 

Another idea to explain HVS not originating from the GC is the origin in a nearby, low-mass galaxy {\it (15)}. Since the escape velocities from those smaller galaxies are smaller as well, it is easier for stars to escape and travel trough the intracluster medium. Some of those neighbouring galaxies also have quite high velocities with respect to our own Galaxy. However, this scenario is also unlikely for US\,708. Although the star might have lived long enough on the main sequence ($\sim10\,{\rm Gyr}$) to travel all the way from a satellite or small neighbouring galaxy, its current state of evolution is quite short compared to its total lifetime (only about $0.1\%$). Furthermore, only about $2\%$ of all main sequence stars undergo an EHB phase. This means that for each single HVS sdO coming from the intracluster medium there should be about $50\,000$ HVS main sequence stars travelling through our Galactic halo. However, only a few tens of them have been reported so far. While faint, high proper motion objects are still not easy to identify, it is very easy to discover stars with high RVs in big survey like SDSS or RAVE. Palladino et al. {\it (15)} list more exotic mechanisms like interactions in globular clusters, with intermediate mass black holes or between galaxies {(\it 54)}. { In addition to that, combinations of several scenarios are imaginable. If for example a hierarchical triple system would be disrupted by the SMBH and one component of the ejected binary would explode as core-collapse SN, the trajectory of the surviving companion would not point back to the GC.} However, we also regard all those scenarios as very unlikely to explain the object presented here. 

\subsection*{Estimating supernova rates}

Another sanity check for our scenario is to provide a rough estimate of the double-detonation SN\,Ia rates we would expect based on our observations and binary population synthesis models. So far US\,708 is unique among the known He-sdO stars and this estimate is based on very small number statistics. The star  was drawn from a sample of hot subdwarfs selected from SDSS. The full sample contains $1369$ hot subdwarfs in total, $262$ of them or roughly $20\%$ are He-sdOs {\it (55)}. Binary population synthesis models by Han et al. {\it (9)} predict a birthrate of $5\times10^{-2}\,{\rm yr^{-1}}$ for hot subdwarfs in general and therefore $1\times10^{-2}\,{\rm yr^{-1}}$ for He-sdOs. One of the observed He-sdOs (US\,708) might be an ejected donor remnant ($\sim0.4\%$). This translates into a double-detonation SN rate of roughly $4\times10^{-5}\,{\rm yr^{-1}}$. This has to be regarded as lower limit only, because a few He-sdOs with smaller RVs, but rather high proper motions might still be hidden in our sample. 

The predicted rates of double-detonation SN\,Ia are around $3\times10^{-4}\,{\rm yr^{-1}}$ and the observed rates of all types of SN\,Ia around $3\times10^{-3}\,{\rm yr^{-1}}$ {\it (56)}. Since all those numbers have quite significant uncertainties, they are regarded as broadly consistent. The most important point for this sanity check is, that our estimates from observations do not deviate from the predicted or observed rates by orders of magnitude.

Yu \& Tremaine {\it (57)} calculated the ejection rates of hypervelocity stars expected from interactions with the Galactic centre black hole (or a binary black hole in the GC). The rates are of the order of $\sim10^{-5}\,{\rm yr^{-1}}$ to $\sim10^{-4}\,{\rm yr^{-1}}$. However, these numbers correspond to the simplest case, the ejection of single main sequence stars. Since stars as peculiar as US\,708 and their progenitors are very rare compared to normal main sequence stars, the close encounter and ejection rates of such stars have to be orders of magnitude smaller. It is therefore very unlikely to find one He-sdO along with the about $20$ other hypervelocity stars assuming that they are all accelerated in the GC.

\vspace{1cm}

\includegraphics[width=14cm]{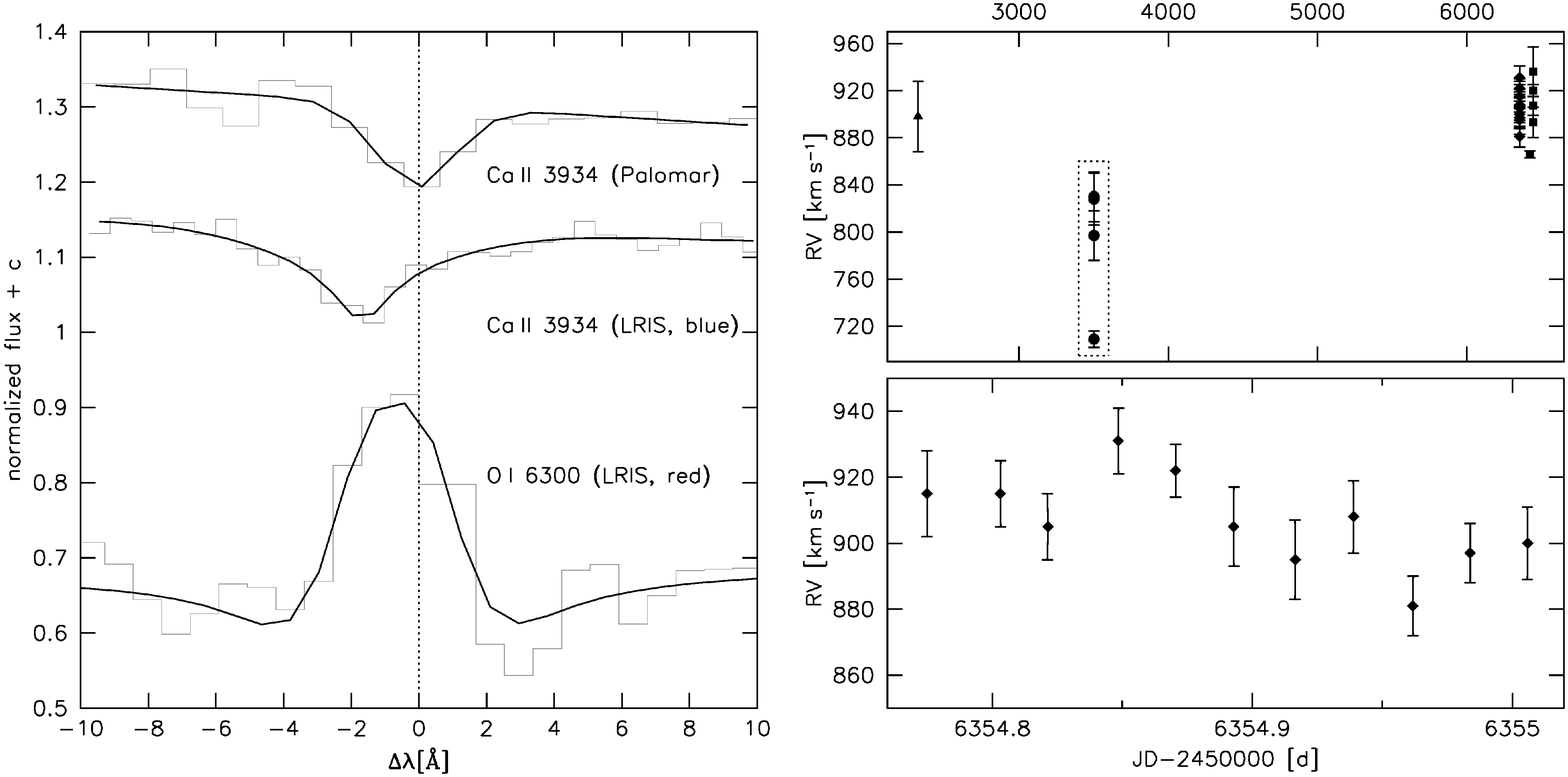}

{ Fig.~S1: Revised radial velocity.} {\it Left panel:} Interstellar Ca\,{\sc ii} line of US\,708 in the Palomar (upper plot) and blue-channel LRIS spectra (middle plot). In contrast to the Palomar spectrum, the LRIS spectrum is significantly blue-shifted. Nightsky emission line of O\,{\sc i} in the LRIS red-channel spectrum (lower panel). The blue-shift is smaller than in the red-channel spectrum. {\it Right panel:} Radial velocities of US\,708 plotted against Julian date. Upper panel: SDSS (triangle), LRIS blue- and red-channel uncorrected (grey circles), LRIS blue- and red-channel corrected (black circles), ESI (diamonds), Palomar (squares). The dotted box marks the LRIS RVs, which are affected by systematics. Lower panel: Close-up of the ESI RVs taken within one night. 

\includegraphics[width=10cm]{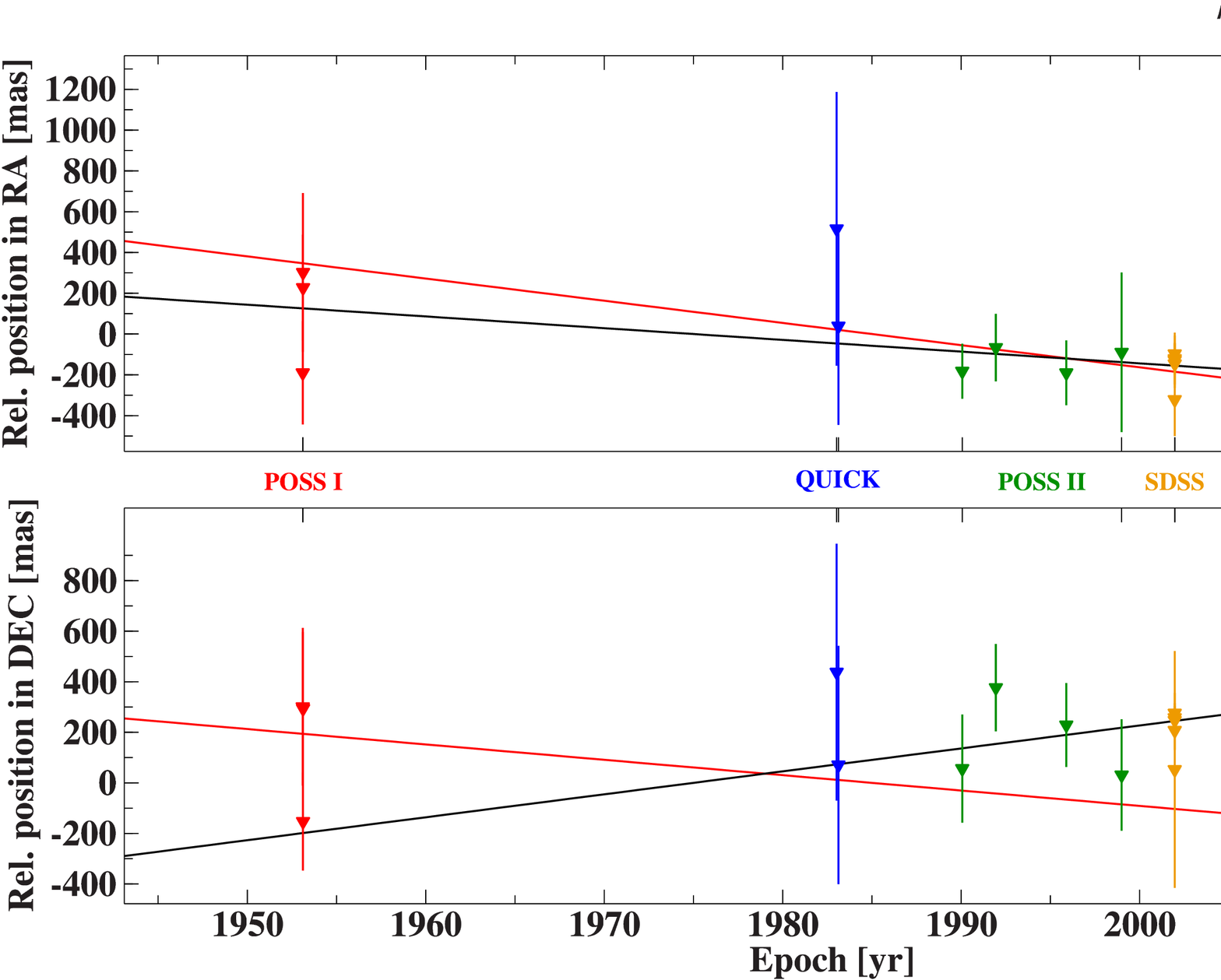}

{ Fig.~S2: Proper motion of US\,708.} Relative positions of US\,708 in right ascension (upper panel) and declination (lower panel) plotted against time. The POSS\,I, QUICK and POSS\,II positions are measured from scanned photographic plates provided by the Digitised Sky Survey. SDSS and PanSTARRS positions are measured from CCD images. The black solid lines mark the best fits, from which we derive the proper motion components. The solid red lines mark the proper motion components required for the star to originate from the Galactic centre.

\includegraphics[width=14cm]{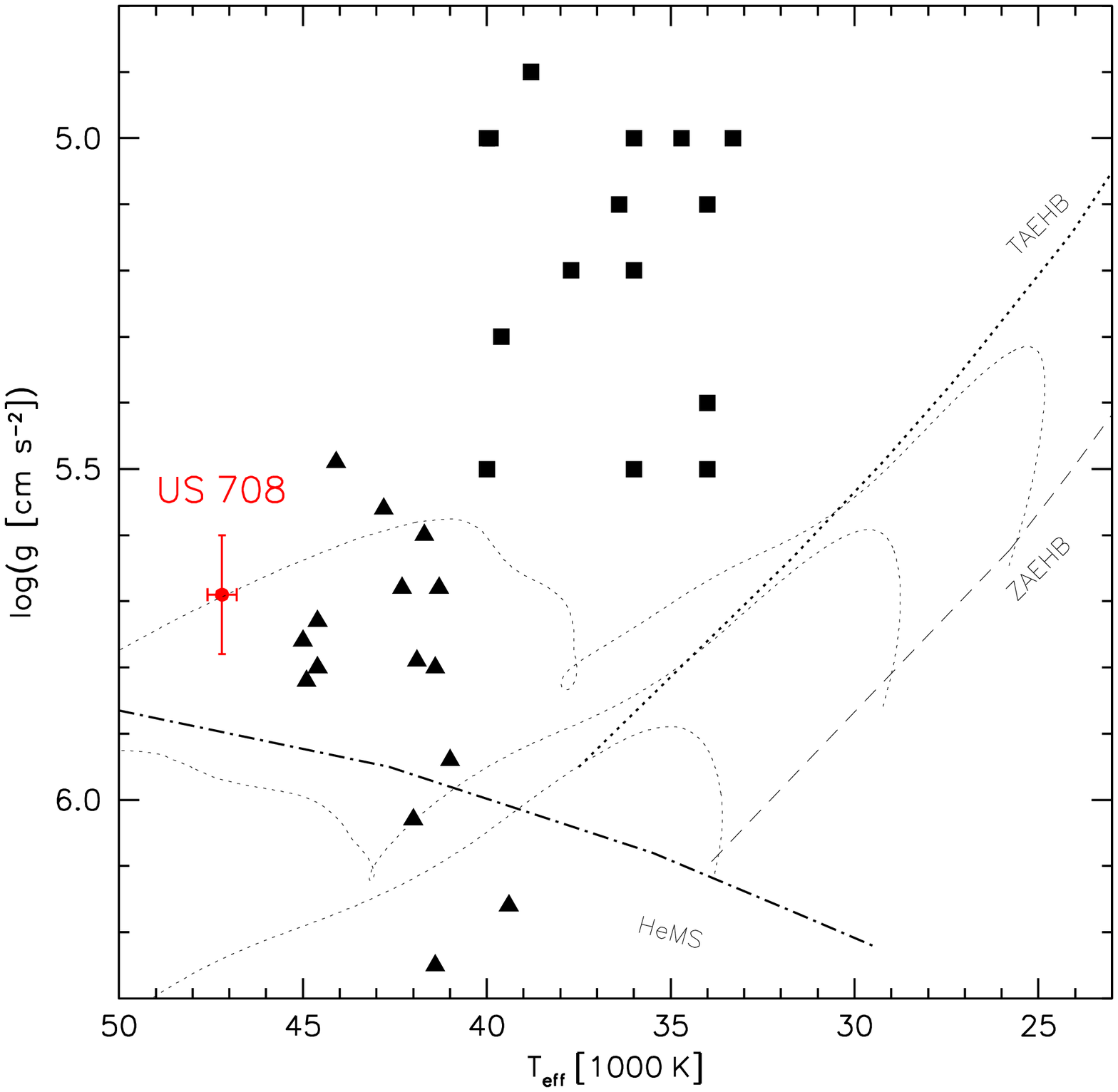}

{ Fig.~S3: Evolutionary status of US\,708.} $T_{\rm eff}-\log{g}$ diagram. Evolutionary tracks (solar metallicity) of core helium-burning stars with a mass of $0.45\,M_{\rm \odot}$ and different masses of their hydrogen envelopes (for bottom to top, $0.0\,M_{\rm \odot}$, $0.001\,M_{\rm \odot}$, $0.005\,M_{\rm \odot}$) are plotted {\it (6)}. The positions of both the Zero Age and the Terminal Age Extended Horizontal Branch (ZAEHB, TAEHB) are indicated as well as the helium main sequence (He-MS). The filled black symbols mark known He-sdBs {\it (37,38)} (squares) and He-sdOs{\it (39)} (triangles) from the literature.

\includegraphics[width=14cm]{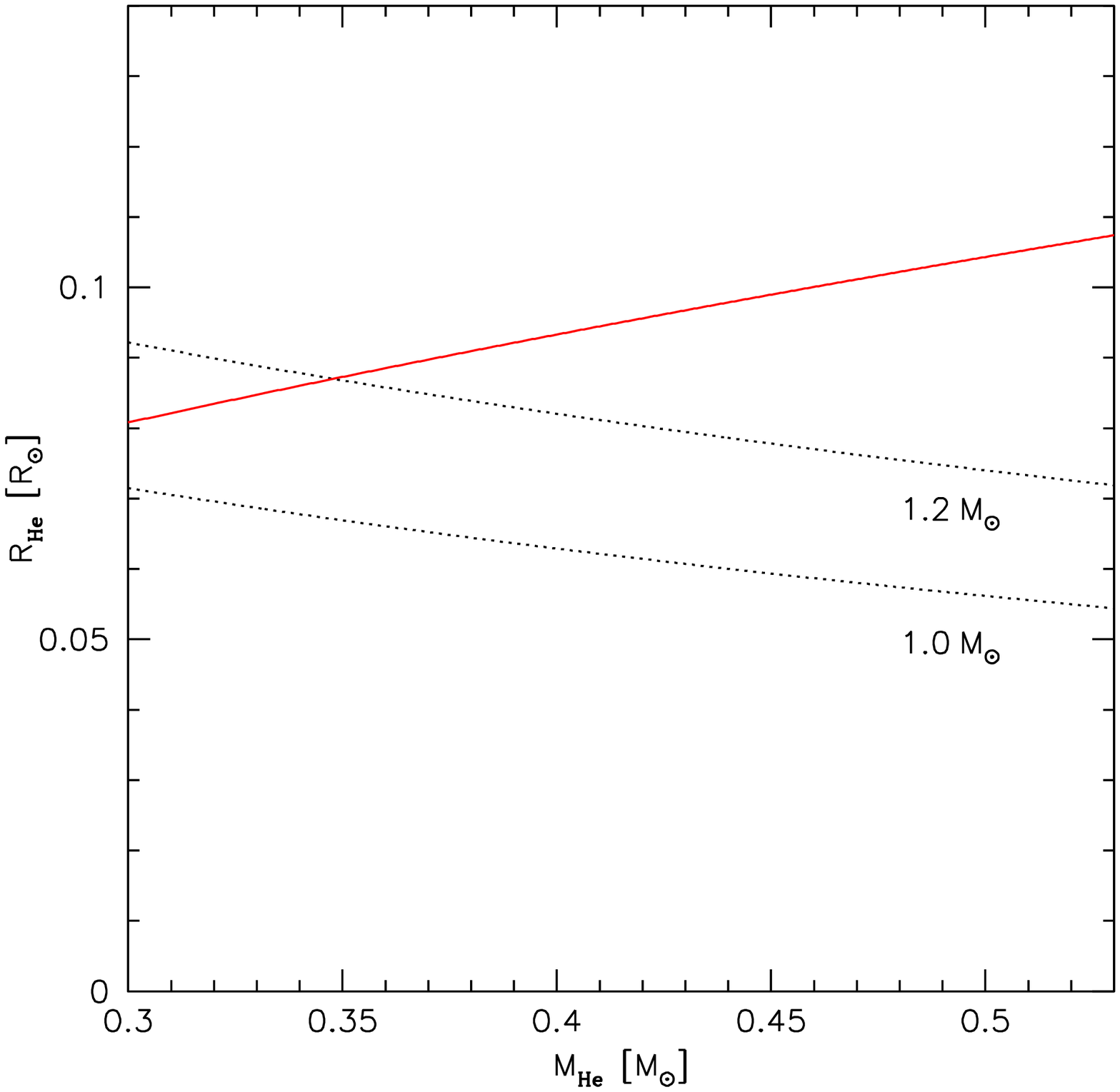}

{ Fig.~S4: Mass-radius relation of the compact He-star.} The dotted black lines mark the Roche radii for WD companion masses of $1.0\,M_{\rm \odot}$ and $1.2\,M_{\rm \odot}$. The red solid line marks the He-star radius assuming $\log{g}=6.1$.

\includegraphics[width=10cm,angle=-90]{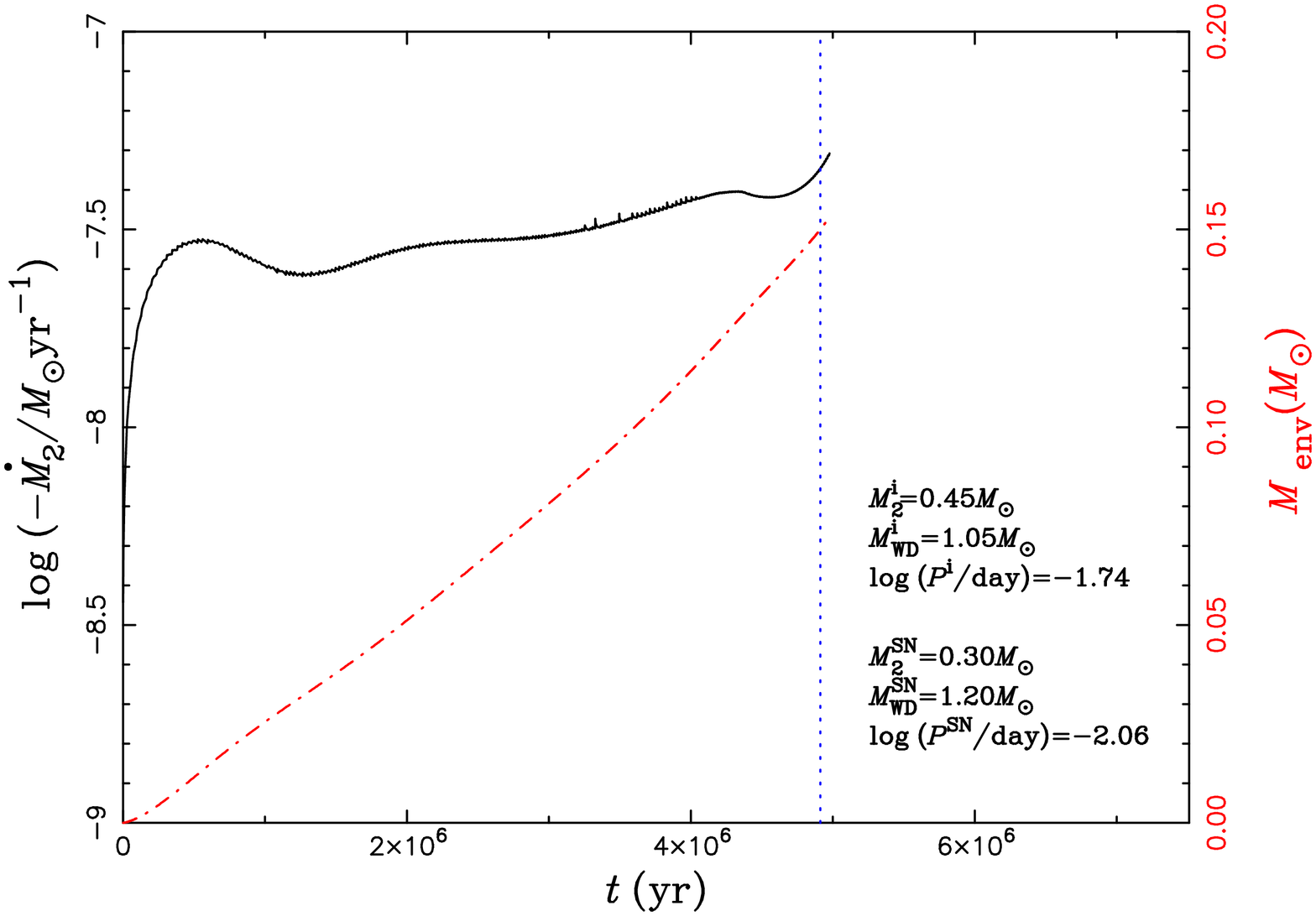}

{ Fig.~S5: Mass-transfer rate.} The solid and dash-dotted curves show the mass-transfer rate and the mass of the WD envelope (He shell) varying with time after the He star fills its Roche lobe, respectively. The dotted vertical line indicates the position where the double-detonation may happen (the mass of the He shell increases to $\sim0.15\,M_{\odot}$). The initial binary parameters and the parameters at the moment of the SN explosion are also given.

\includegraphics[width=10cm,angle=-90]{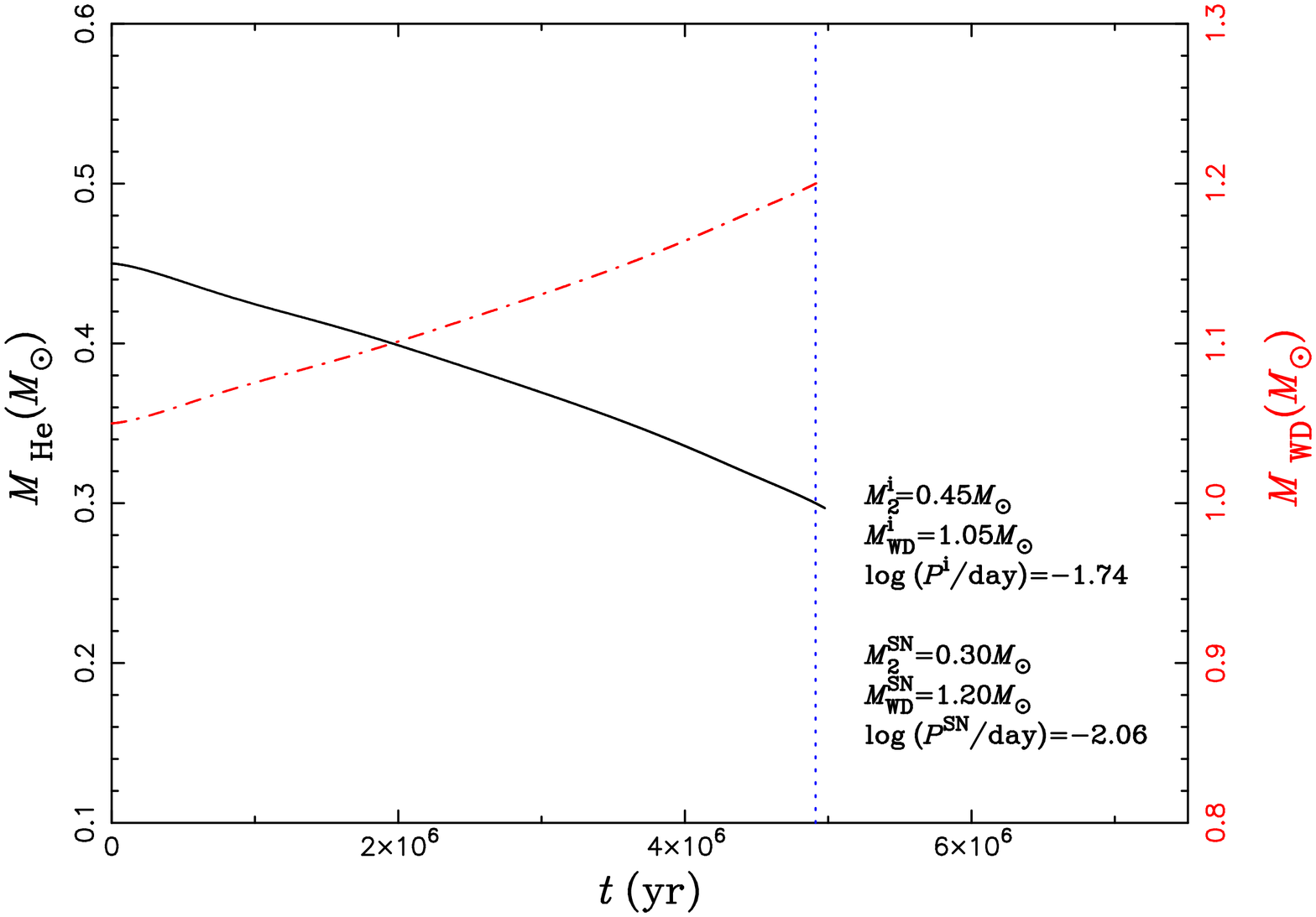}

{ Fig.~S6: Change of component masses.} Change of He-star (solid line) and WD mass (dash-dotted line) with time. The dotted vertical line indicates the position where the double-detonation may happen (the mass of the He shell increases to $\sim0.15\,M_{\odot}$). The initial binary parameters and the parameters at the moment of the SN explosion are also given.

\includegraphics[width=10cm,angle=-90]{extended_fig7.ps}

{ Fig.~S7: Evolution of orbital parameters.} Time evolution of the radial velocity semiamplitude (solid line) and the orbital period (dash-dotted line) of the binary. The dotted vertical line indicates the position where the double-detonation may happen (the mass of the He shell increases to $\sim0.15\,M_{\odot}$). The initial binary parameters and the parameters at the moment of the SN explosion are also given.

\end{document}